\documentclass[12pt]{article}
\renewcommand{\theequation}{\thesection.\arabic{equation}}
\newcounter{subequation}[equation]
\makeatletter
 
\expandafter\let\expandafter\reset@font\csname reset@font\endcsname
 
\def\subeqnarray{\arraycolsep1pt
    \def\@eqnnum\stepcounter##1{\stepcounter{subequation}%
        {\reset@font\rm(\theequation\alph{subequation})}}
\jot5mm     \eqnarray}

\makeatother\newcommand{\newsection}[1]{\vspace{10mm}
\pagebreak[3]\addtocounter{section}{1}\setcounter{equation}{0}
\setcounter{subsection}{0}\setcounter{footnote}{0}
 
\begin{flushleft}{\Large\bf \thesection. #1}
\end{flushleft}\nopagebreak\medskip\nopagebreak}

\def\tr{\mathop{\hbox{\rm tr}}\nolimits}
\def\Or{\hbox{\rm O}}
\def\rmd{\hbox{\rm d}}
\def\rme{\hbox{\rm e}}
\def\rmi{\hbox{\rm i}}
\def\eref#1{(\ref{#1})}
\def\be{\begin{displaymath}}
\def\ee{\end{displaymath}}
\def\beq{\begin{equation}}
\def\eeq{\end{equation}}
\def\bea{\begin{eqnarray}}
\def\eea{\end{eqnarray}}
\def\l{\lambda}

\def\dd{\partial}

\def\vac{\left|0\right>}
\def\ket#1{\left|#1\right>}
\def\dfrac#1#2{\frac{\displaystyle #1}{\displaystyle #2}}
\def\one#1{#1^{\raise5pt\hbox{$\scriptstyle\!\!\!\!1$}}\,{}}
\def\two#1{#1^{\raise5pt\hbox{$\scriptstyle\!\!\!\!2$}}\,{}}
\def\re{\mathop{\hbox{\rm Re}}\nolimits}
\def\id{\mathop{\hbox{\rm id}}\nolimits}
\def\binom#1#2{\left(#1 \atop #2\right)}
\def\tilde{\widetilde}
\makeatletter
\def\binrel@#1{\begingroup
  \setboxz@h{\thinmuskip0mu
    \medmuskip\m@ne mu\thickmuskip\@ne mu
    \setbox\tw@\hbox{$#1\m@th$}\kern-\wd\tw@
    ${}#1{}\m@th$}%
  \edef\@tempa{\endgroup\let\noexpand\binrel@@
    \ifdim\wdz@<\z@ \mathbin
    \else\ifdim\wdz@>\z@ \mathrel
    \else \relax\fi\fi}%
  \@tempa
}
\let\binrel@@\relax
\def\overset#1#2{\binrel@{#2}%
  \binrel@@{\mathop{\kern\z@#2}\limits^{#1}}}
\def\underset#1#2{\binrel@{#2}%
  \binrel@@{\mathop{\kern\z@#2}\limits_{#1}}}
\makeatother
\newfont{\bbd}{msbm10 scaled\magstep1}
\def\C{\hbox{\bbd C}}
\def\MM{\hbox{\bbd M}}
\def\R{\hbox{\bbd R}}

\def\M{{\mathcal M}}

\def\PP{{\mathcal P}}
\def\LL{{\mathcal L}}
\def\MM{{\mathcal M}}
\def\QQ{{\mathcal Q}}
\def\RR{{\mathcal R}}
\def\WW{{\mathcal W}}
\def\om{\omega}
\def\Cx{\C[\vec x]}
\def\Cy{\C[\vec y]}
\newfont{\frak}{eufm10 scaled\magstep1}
\def\JPA{J.\ Phys.\ A: Math.\ Gen.\ }

\def\LMP{Lett.\ Math.\ Phys.\ }
 
\def\bt{B\"ack\-lund tran\-s\-for\-ma\-ti\-on}
\begin{document}
\begin{flushright}
\sf LPENSL-TH-16/99 \\
\sf solv-int/9908002
\end{flushright}
\vskip1cm
\begin{center}\LARGE\bf
Quantum B\"{a}cklund transformation \\
for the integrable DST model
\end{center}
\vskip1cm
\begin{center}
V B Kuznetsov,\dag\footnote[1]{E-mail: {\tt vadim@amsta.leeds.ac.uk}}
\ M Salerno\ddag\ 
and E K Sklyanin\S\footnote[2]{On leave from: Steklov Mathematical Institute 
at St.~Petersburg, Fontanka 27, St.~Petersburg 191011, Russia. 
E-mail: {\tt sklyanin@pdmi.ras.ru}}
\vskip0.4cm
\dag Department of Applied Mathematics, 
University of Leeds, Leeds LS2 9JT, UK\\
\ddag Department of Theoretical Physics, University of Salerno 84081,
Salerno, Italy \\
\S\ Laboratoire de Physique%
\footnote[3]{UMR 5672 du CNRS et de l'ENS Lyon}, 
Groupe de Physique Th\'eorique, ENS Lyon,
46 all\'ee d'Italie, 69364 Lyon 07, France
\end{center}
\vskip2cm
\begin{center}
\bf Abstract
\end{center}
For the integrable case of the discrete self-trapping (DST) model we
construct a \bt. The dual Lax matrix and
the corresponding dual \bt\ are also found and studied.
The quantum analog of the \bt\ ($Q$-operator) is constructed
as the trace of a monodromy matrix with an infinite-dimensional
auxiliary space. We present the $Q$-operator as an explicit
integral operator as well as describe its action on the monomial basis.
As a result we obtain a family of integral equations for
multivariable polynomial eigenfunctions of the quantum
integrable DST model. These eigenfunctions are special
functions of the Heun class which is beyond the hypergeometric
class. The found integral equations are new and they shall
provide a basis for efficient analytical and numerical
studies of such complicated functions.
\vskip1cm
\noindent 31 July 1999
\vskip1cm
\noindent Submitted to {\it\JPA}
\newpage

\newsection{Introduction}
\setcounter{equation}{0}
\label{sec:intro}

The discrete self--trapping (DST) equation was introduced
by Eilbeck, Scott and Lomdhal \cite{ELS85} to model  
the nonlinear dynamics of small molecules, such as ammonia,
acetylene, benzene, as well as large molecules, such as 
acetanilide. In simple terms, it consists of a set of $n$ 
nondissipative anharmonic oscillators coupled through 
dis\-per\-sive interactions. 
Due to the nonlinearity this system can have  
complicated dynamical be\-ha\-viour going from quasiperiodic motion 
to chaos \cite{chaos1,chaos2}.
The DST equation is also found in connection with physical 
problems in different areas such as stabilization of high frequency 
vibrations in biomolecular dynamics \cite{davydov},  
arrays of coupled nonlinear waveguides in nonlinear optics \cite{fs90},
quasiparticle motion on a dimer \cite{kc86}.
In the case of two degrees of freedom $n=2$  (DST dimer) 
the system is integrable having, besides the Hamiltonian (energy),
another conserved quantity, the norm 
(number of particles in the quantum case). 
The integrability properties of the classical and quantum 
DST dimer  were studied in detail by several methods such as
the number state method \cite{se86}, the algebraic Bethe ansatz \cite{esks},
and the method of separation of variables \cite{EKS93}.
For more than two  degrees of freedoms 
an integrable case of the DST system was found and studied in \cite{CJK93}.
This integrable case is close to the Toda lattice and coincides 
for $n=2$ with the usual DST dimer.

The quantum Hamiltonian $H$ of the integrable DST 
model contains $(n+1)$ parameters 
$c_1,\ldots,c_n,b$ and is defined as a second-order 
differential operator (here $\dd_i\equiv\dd/\dd x_i$)
\beq
H=\sum_{i=1}^n \left(\frac12x_i^2\dd_i^2+(c_i+\frac12)x_i\dd_i
+bx_{i+1}\dd_i\right)
\label{eq:def-H}
\eeq
acting in the space $\Cx$ of polynomials of $n$ variables
$\{x_1,\ldots x_n\}\equiv\vec x$. 
In \eref{eq:def-H} and other similar formulas we always
assume the periodic boundary conditions $x_{n+1}\equiv x_1$.

The Hamiltonian $H$ obviously commutes with the number-of-particles operator
$N$
\beq
N=\sum_{i=1}^n x_i\dd_i.
\label{eq:def-N}
\eeq

As shown in section \ref{sec:quant}, $H$ and $N$ can be included in a 
commutative ring of differential operators generated by a basis of $n$
operators, the fact allowing to claim the quantum integrability of the system.

The multiplication operators $x_i$ 
and the respective differentiations $\dd_i$ can be considered as generators
of a Heisenberg algebra (creation/annihilation operators).
There exists a well-known scalar product
on $\Cx$ (holomorphic representation)
such that $x_i$ and $\dd_i$ become mutually adjoint $\dd_i^\dagger=x_i$.
The corresponding Hamiltonian $H$ is self-adjoint, however, only in the
dimer case $n=2$. In the general $n>1$ case, no involution rendering $H$
self-adjoint is known. 
The Hilbert space structure is, however, quite irrelevant for the kind of
problems we are interested in and will be completely ignored throughout the
paper. 

The DST-chain can be considered as a degenerate case of the 
Heisenberg magnetic chain, though not as degenerate as Toda lattice.
It makes the DST-chain a good tool for studying various techniques
applicable to integrable models since it requires more effort than Toda lattice
but still is simpler than the generic magnetic chain.

The main purpose of the present paper is to construct an analog of Baxter's
$Q$-operator \cite{Bax82} for the integrable DST model. The $Q$-operator $Q_\l$
by definition shares the set of eigenvectors with the Hamiltonians $H_i$,
and its eigenvalues are polynomials in $\l$
satisfying a finite-difference equation known as {\it Baxter} or
{\it separation equation}.
As was shown in \cite{PG92} on the example of the periodic Toda lattice,
the similarity transformation
${\mathcal O}\mapsto Q_\l{\mathcal O}Q_\l^{-1}$
turns in the classical limit
into a classical B\"acklund transformation that is a one-parametric
family of canonical transformations preserving the commuting Hamiltonians.
Later, in \cite{KS5} for the classical
B\"acklund transfomations the property of {\it spectrality} was described
which is the classical counterpart of the separation equation for the
eigenvalues of $Q_\l$.
In the present paper we follow the approach
of \cite{KS5} studying first the classical case and paying the special
attention to the {spectrality} property of the corresponding B\"acklund
transformation.

Our main result (see sections 4--7) is the following integral equation
\beq
 \int_\gamma\rmd\xi_1\ldots \int_\gamma\rmd\xi_n\,
 \left[\,\prod_{j=1}^n \frac{\rmi}{2\pi}\Gamma(\l+1-c_j)
 \rme^{-\xi_j}(-\xi_j)^{c_j-\l-1}\right]
\,\psi(\ldots,y_k\xi_k+by_{k+1},\ldots)
\label{new1}
\eeq
\beq
=q(\l)\psi(x_1,\ldots,x_n),\qquad q(\l)\in\C[\l],
\label{new2}
\eeq
for the polynomial eigenfunctions $\psi\in\Cx$ of the Hamiltonian
\eref{eq:def-H}
\beq
H\psi(x_1,\ldots,x_n)=h\psi(x_1,\ldots,x_n).
\eeq

The structure of the paper is the following. In section 2
we consider the
classical version of the integrable DST-chain and describe its relation
to the Toda lattice and the isotropic Heisenberg magnetic chain.
Our construction of \bt\ generalizes
well known results for the Toda lattice. Following \cite{KS5}, we also study
the dual Lax matrix and the corresponding dual \bt\ in section 3.

In section 4 we discuss the quantization of the integrable DST
model and present a list of properties of Baxter's $Q$-operator.
In section 5, following the approach of \cite{BLZ},
we construct a $Q$-operator $Q_\l$ for the quantum DST chain as
the trace of a monodromy matrix with an infinite-dimensional auxiliary space.
In the spirit of \cite{PG92}, we consider $Q_\l$ as
an integral operator in $\Cx$ and find in section 6
its kernel and contour of integration.
In the same section
we study analiticity properties of $Q_\l$ in the parameter $\l$,
prove that its matrix elements in the monomial basis are polynomials in $\l$
and give explicit formulas for its action on polynomials.
We consider in details the simplest $n=1$ case where
the $Q$-operator provides an integral representation for
classical orthogonal polynomials (Charlier polynomials).
In section 7 we prove that $Q_\l$ satisfies a
finite-difference equation in the parameter $\l$.
In the last, 8$th$ section we discuss 
possible generalizations and applications of our results.

\newsection{Classical case}
\setcounter{equation}{0}
\label{sec:classic}

In this section we consider the classical integrable DST-chain \cite{CJK93}.
The model is described in terms of $n$ pairs of canonical variables
$(X_i,x_i)$, $i=1,\ldots,n$ 
\beq
 \{X_i,X_j\}=\{x_i,x_j\}=0, \qquad \{X_i,x_j\}=\delta_{ij}
\label{eq:pb-xX}
\eeq
(the periodicity convention $x_{i+n}\equiv x_i$, $X_{i+n}\equiv X_i$
is always assumed for the indices of $x_i$ and $X_i$).

The canonical momenta $X_i$ replace in the classical case the differential
operators $\dd_i$. As mentioned before, in the quantum case we do not make any 
assumptions about self-adjointness of the observables. Respectively, we
allow the classical variables $(X_i,x_i)$ to be complex.

To construct $n$ commuting Hamiltonians we introduce the Lax matrix $L(u)$
(monodromy matrix) as a product of $n$ local Lax matrices $\ell_i(u)$
\beq
 L(u)=\ell_n(u)\ldots\ell_2(u)\ell_1(u)
\label{eq:monodromy}
\eeq
\beq
 \ell_i(u;x_i,X_i)=\left(\begin{array}{cc}
       u-c_i-x_iX_i & b x_i \\ -X_i & b \end{array}\right)
\label{eq:def-ell}
\eeq
where $b,c_i\in\C$ are 
parameters of the model, and $u$ is the so-called
spectral parameter of the Lax matrix.

Denoting by $\id_2$ the unit $2\times2$ matrix and
introducing notations for the tensor products
$\one\ell\equiv\ell\otimes\id_2$,
$\two\ell\equiv\id_2\otimes\ell$ one establishes the
$r$-matrix identity \cite{FTbook}
\beq
 \{\one\ell_i(u_1),\two\ell_j(u_2)\}
   =[r_{12}(u_1-u_2),\one\ell_i(u_1)\two\ell_j(u_2)]\delta_{ij}, \qquad
   r_{12}(u)=-\frac{1}{u}\PP_{12}
\label{eq:pb-ll}
\eeq
where $\PP_{12}$ is the permutation operator in $\C^2\otimes\C^2$.
{}From \eref{eq:pb-ll} the corresponding identity for the monodromy
matrix
\beq
 \{\one L(u_1),\two L(u_2)\}
   =[r_{12}(u_1-u_2),\one L(u_1)\two L(u_2)]
\label{eq:pb-LL}
\eeq
is derived in the standard way \cite{FTbook} which, in turn, ensures
the commutativity of the spectral invariants $t(u)$ and $d(u)$
of the matrix $L(u)$
defined as coefficients of its characteristic polynomial
\beq
\det(v-L(u))=v^2-t(u)v+d(u).
\eeq

Since $\det\ell_i(u)=b(u-c_i)$ the determinant 
$d(u)\equiv\det L(u)=\prod_{i=1}^n b(u-c_i)$ is scalar, and the only nontrivial
spectral invariant is the trace $t(u)$
\beq
t(u)\equiv\tr L(u)=L_{11}(u)+L_{22}(u)
\label{eq:def-t}
\eeq
which serves as a
generating function of commuting independent Hamiltonians $H_i$
\beq
t(u)=u^n+\sum_{i=1}^n (-1)^i H_iu^{n-i}.
\label{eq:def-Hi}
\eeq

As a corollary of \eref{eq:pb-LL} we have the commutativity of $t(u)$
\beq
 \{t(u_1),t(u_2)\}=0
\label{eq:comm-t}
\eeq
and, consequently, the commutativity $\{H_i,H_j\}=0$ of the Hamiltonians $H_i$.

A direct calculation shows that
\beq
 H_1=N+\sum_{i=1}^n c_i, \qquad  H_2=\frac12H_1^2-H-\frac12\sum_{i=1}^nc_i^2,
\label{eq:def-H12}
\eeq
where
\beq
N=\sum_{i=1}^n x_iX_i, \qquad
H=\sum_{i=1}^n \left(\frac12x_i^2X_i^2+c_ix_iX_i+bx_{i+1}X_i\right)
\eeq
ensuring that the polynomial ring of commuting Hamiltonians contains the 
number of particles $N$ and the Hamiltonian $H$.

Note that the $r$-matrix $r_{12}(u)$ in \eref{eq:pb-ll} 
is the same as for the isotropic
Heisenberg magnetic chain and Toda lattice \cite{FTbook} which puts these
integrable models into the same class. Indeed, the Toda lattice is 
a degenerate case of the DST chain. To demonstrate it, it is sufficient to
make a constant shift $u\mapsto u+b^{-1}$ of the spectral parameter
in $\ell_i(u)$ given by \eref{eq:def-ell} and take the limit 
\beq
b\rightarrow 0,\qquad x_j=\rme^{q_j}(b^{-1}+p_j)+\Or(b),\qquad
X_j=\rme^{-q_j},
\label{eq:b->0}
\eeq
contracting the `oscillator' algebra $(x_i,X_i,x_iX_i)$ into the
Euclidian Lie algebra $(\rme^{\pm q_i},\allowbreak p_i)$. In the limit
$\ell_i(u)$ turns into the elementary $\ell$-matrix for the
Toda lattice:
\beq
\ell_i(u)\rightarrow\left(\begin{array}{cc}
       u-c_i-p_i & \rme^{q_i} \\ -\rme^{-q_i} & 0
\end{array}\right)
\eeq
(the shifts $c_i$ become irrelevant since they can be absorbed into a
simple canonical transformation $p_i\mapsto p_i-c_i$).
On the other hand, the DST model, in turn, is a degenerate case of the
Heisenberg XXX magnet corresponding to the contraction of the $su(2)$ Lie 
algebra into the oscillator algebra. The DST model holds 
an intermediate place 
between Heisenberg and Toda models.

In the present paper we take the Hamiltonian point of view on \bt\
according to which
the \bt\ $B_\l$ is a one-parameter family of simplectic maps
from the canonical variables $(\vec X,\vec x)$
to the canonical variables $(\vec Y,\vec y)$
possessing certain characteristic properties
(see \cite{KS5} for a detailed discussion).
For Hamiltonian integrable systems allowing a description in terms
of the $r$-matrix algebra \eref{eq:pb-LL} there has recently been found
an algorithmic method for constructing a \bt\ \cite{Skl51,Skl52}.
The method having been described in detail in the cited papers, we present
here only the results.

As in the case of the periodic Toda lattice \cite{PG92,KS5}, it is convenient
to describe the canonical transformation $B_\l$
in terms of the generating function
\beq
 F_\l(\vec y\mid\vec x)=n\l+\sum_{i=1}^n\left(
   \frac{x_i-by_{i+1}}{y_i}
  +(\l-c_i)\ln\frac{b y_{i+1}-x_i}{(\l-c_i)by_i}\right).
\label{eq:def-F}
\eeq

\begin{subeqnarray}
  X_i&=&\dfrac{\dd F_\l}{\dd x_i}=\dfrac{1}{y_i}+\dfrac{\l-c_i}{x_i-b y_{i+1}}
\\
 Y_i&=&-\dfrac{\dd F_\l}{\dd y_i}
=bX_{i-1}+\dfrac{x_i-b y_{i+1}}{y_i}\,X_i.
\label{eq:def-XY}
\end{subeqnarray}

To prove that $B_\l$ preserves the Hamiltonians $H_i$
\beq
   H_i(\vec X,\vec x)=H_i(\vec Y,\vec y)
\label{eq:inv-H}
\eeq
we proceed in the same manner as in \cite{PG92,KS5}
for the periodic Toda lattice. Introducing the matrices 
\beq
 M_i(u)=\left(\begin{array}{cc}
       1 & -by_{i+1} \\ X_i & u-\l-by_{i+1}X_i
       \end{array}\right),
\label{eq:def-M}
\eeq
one then directly verifies the equality
\beq
    M_{i}(u)\ell_i(u;X_i,x_i)=\ell_i(u;Y_i,y_i)M_{i-1}(u)
\label{eq:comm-Ml}
\eeq
from which it follows that $B_\l$ preserves the spectrum of the Lax
matrix $L(u)$
\be
 M_n(u,\l)L(u;\vec X,\vec x)=L(u;\vec Y,\vec y)M_n(u,\l)
\ee
which, in turn, ensures the invariance of $t(u)$ and, therefore,
of $H_i$ \eref{eq:inv-H}.

To formulate the {\it spectrality property} \cite{KS5} of the B\"acklund
transformation we introduce the quantity $\mu$ canonically conjugated,
in a sense, to $\l$
\beq
 \ln\mu=-\frac{\dd F_\l}{\dd\l}=
     \sum_{i=1}^n\ln\frac{(\l-c_i)by_i}{b y_{i+1}-x_i}, \qquad
 \mu=\prod_{i=1}^n\frac{(\l-c_i)by_i}{b y_{i+1}-x_i}.
\label{eq:def-mu}
\eeq

The spectrality of the \bt\ means that the pair $(\l,\mu)$ lies on
the spectral curve of the Lax matrix
\beq
\det(\mu-L(\l))=0.
\label{eq:mu-la}
\eeq

To prove it, we again follow \cite{KS5}. We observe that for $u=\l$ the
matrix
$M_i(u)$ degenerates
\beq
 M_i(\l)=\left(\begin{array}{cc}
       1 & -by_{i+1} \\ X_i & -by_{i+1}X_i
       \end{array}\right)
       =\left(\begin{array}{c}
        1 \\ X_i
       \end{array}\right)
       \begin{array}{cc}
       (1 & -by_{i+1}) \\ \phantom{(1} & \phantom{s_i)}
       \end{array},
\label{eq:M(la)}
\eeq
and its null-vector $\om_i$ can be found explicitly:
\beq
 M_i(\l)\om_i=0, \qquad
     \omega_i=\left(\begin{array}{c} b y_{i+1} \\ 1 \end{array}\right).
\label{eq:Mom}
\eeq

Noting then the identity
\beq
 \ell_i(\l)\om_{i-1}=\frac{(\l-c_i)b y_i}{b y_{i+1}-x_i}\om_{i},
\label{eq:ell-om}
\eeq
we conclude that
\beq
 L(\l)\om_n=\mu\om_n,
\label{eq:Lom}
\eeq
whence \eref{eq:mu-la} follows immediately.

The commutativity $B_{\l_1}\circ B_{\l_2}=B_{\l_2}\circ B_{\l_1}$ is an
immediate consequence of the invariance of Hamiltonians and their
completeness, see \cite{KS5}.

Note that $M_i^{-1}(u)$ and $\ell_i(u)$ have, as functions of $u$,
essentially the same structure, up to a shift of $u$ and a scalar factor.
The fact is by no means a coincidence, see \cite{Skl52}
for a detailed explanation.

\newsection{Dual Lax matrix}
\setcounter{equation}{0}
\label{sec:dual}

We conclude the study of the classical case
with presenting the {\it dual} Lax matrix
and the dual \bt\ for the DST model. In \cite{CJK93}
two different Lax matrices were found for the integrable
DST system, the $2\times2$ Lax matrix $L(u)$ and also the $n\times n$
Lax matrix. This bigger Lax matrix did not contain a spectral
parameter. Here we present an $n\times n$ Lax matrix $\LL(v)$ containing
a spectral parameter $v$ which is dual to $L(u)$ in the sense
that the corresponding spectral curves are equivalent up to interchanging
the spectral parameters $u$ and $v$ 
\beq
(b^n-v)\det\left(u-\LL(v)\right)=\det(v-L(u)).
\label{eq:duality-L}
\eeq

To produce the dual Lax matrix $\LL(v)$ we take an eigenvector $\theta_1(u)$
of $L(u)$ corresponding to the eigenvalue $v$ (for brevity, we will
not mark the dependence on $u$ in $\theta$)
\beq
 L(u)\theta_1=v\theta_1
\label{eq:def-theta-1}
\eeq
and define by induction $\theta_i$ as
\beq
 \theta_{i+1}=\ell_i(u)\theta_i, \qquad i=1,\ldots,n.
\label{eq:def-theta-i}
\eeq

{}From \eref{eq:def-theta-1} it follows that $\theta_{n+1}=v\theta_1$. The
function $\theta_i(u)$, when properly normalized,
is called {\it Baker-Akhiezer  function}.
Denoting the components  of the vector $\theta_i$ as $\varphi_i$ and $\psi_i$
we write down \eref{eq:def-theta-i} explicitly as
\beq
\left(\begin{array}{cc} \varphi_{i+1} \\ \psi_{i+1} \end{array}\right) =
\left(\begin{array}{cc} u-c_i-x_iX_i & b x_i \\ -X_i & b \end{array}\right)
 \left(\begin{array}{cc} \varphi_{i} \\ \psi_{i} \end{array}\right).
\eeq

Then, splitting the components and taking into account the quasiperiodicity
condition $\theta_{n+1}=v\theta_1$ we arrive to the following linear
equations for $\varphi_i$ and $\psi_i$:

\begin{subeqnarray}
 u\varphi_i&=&\varphi_{i+1}+(c_i+x_iX_i)\varphi_i-bx_i\psi_i, \qquad
 i=1,\ldots,n-1 \\
  u\varphi_n&=&v\varphi_{1}+(c_n+x_nX_n)\varphi_n-bx_n\psi_n,
\label{eq:eqs-phi}
\end{subeqnarray}
\begin{subeqnarray}
  \psi_{i+1}&=&-X_{i}\varphi_i+b\psi_i, \qquad i=1,\ldots,n-1 \\
 v\psi_1&=&-X_n\varphi_n+b\psi_n.
\label{eq:eqs-psi12}
\end{subeqnarray}

Eliminating $\psi_i$ we can write down the linear problem for the
vector $\Phi$  with the components $\varphi_i$ in the matrix form:
\beq
\LL(v)\Phi=u\Phi, \qquad 
 \Phi=\left(\begin{array}{c} \varphi_1 
\\ \ldots \\ \varphi_n \end{array}\right)
\eeq 
where the matrix $\LL(v)$ defined as
\bea
\LL(v)=&(v-b^n)^{-1}\sum_{j,k=1}^n b^{n+j-k}x_jX_kE_{jk}
           +vE_{n1} \nonumber \\
       & +\sum_{j\geq k}b^{j-k}x_jX_kE_{jk}
         +\sum_{j=1}^nc_jE_{jj}
         +\sum_{j=1}^{n-1}E_{j,j+1}
\label{eq:def-LL}
\eea
is the dual Lax matrix we were looking for.
Here $E_{jk}$ is the $n\times n$ matrix with the only non-zero entry
$(E_{jk})_{jk}=1$. The proof of the identity \eref{eq:duality-L} is an
exercise which we leave to the reader. 
For the case $b=1$ and $v=-1$ the dual Lax matrix for the DST model was first 
found in \cite{CJK93}. 
For examples of Lax matrices duality in other
integrable models see \cite{dual}.

The \bt\ ${\mathcal B}_\mu$
corresponding to the dual Lax operator $\LL(v)$ is given by
the same equations \eref{eq:def-XY} and \eref{eq:def-mu}.
The important difference, however, is that now $\mu$ is
a free numerical parameter of \bt\ whereas $\l$ becomes a dynamical 
variable determined from the equation \eref{eq:def-mu}. 
The equality \eref{eq:def-mu} is now reinterpreted as the equation
defining the variable $\l$.  The generating function of ${\mathcal B}_\mu$
is the Legendre transform of $F_\l(\vec y\mid \vec x)$ with respect to $\l$.

The properties of the dual \bt\ ${\mathcal B}_\mu$ are proved in the same 
manner as those of $B_\l$ (see also \cite{KS5} for the Toda lattice case).
For the proof we need a matrix $\MM(v)$ playing for $\LL(v)$ the same
role that $M_n(u)$ played  for $L(u)$.

Let $\tilde\theta_i$ be defined as
$\tilde\theta_i=M_{i-1}\theta_i$. From \eref{eq:comm-Ml}
it follows that $\tilde\theta_i$ is a Baker-Akhiezer function for 
$\ell_i(u;Y_i,y_i)$. The first component of the equality 
$\tilde\theta_i=M_{i-1}\theta_i$ reads 
$\tilde\varphi_i=\varphi_i-by_i\psi_i$. Substituting $\psi_i$ 
from the solution 
to the system \eref{eq:eqs-psi12} we obtain 
the correspondence $\tilde\Phi=\MM(v)\Phi$ with
the matrix $\MM(v)$ defined as
\beq
 \MM(v)=(v-b^n)^{-1}\sum_{j,k=1}^n b^{n+j-k}y_jX_k E_{jk}
   +\sum_{j>k}b^{j-k}y_jX_k E_{jk}
   + \sum_{j=1}^n E_{jj}.
\label{eq:def-MM}
\eeq

The invariance of the spectrum of $\LL(v)$ follows from the identity
\beq
 \MM(v)\LL(v;\vec X,\vec x)=\LL(v;\vec Y,\vec y)\MM(v).
\label{eq:ML=LM dual}
\eeq

The spectrality is expressed as the identity
\beq
\det\bigl(\l-\LL(\mu)\bigr)=0.
\label{eq:dual-sp}
\eeq

To prove \eref{eq:dual-sp} it is sufficient to notice that the matrix
$\MM(v)$ degenerates as $v=\mu$
\beq
 \det\MM(\mu)=0, 
\eeq
and the corresponding null-vector $\Omega$ defined by the recurrence relation
\beq
 \frac{\Omega_{i+1}}{\Omega_i}=\frac{b(c_i-\l)y_{i+1}}{x_i-by_{i+1}},
\qquad i=1,\ldots,n-1
\eeq
is, by virtue of  \eref{eq:ML=LM dual}, also an eigenvector of
$\LL(\mu)$ corresponding to the eigenvalue $\l$
\beq
  \LL(\mu)\Omega=\l\Omega.
\eeq

Since the Toda lattice is a degenerate case of the DST model, 
the $n\times n$ Lax matrix for the Toda lattice
can be obtained, as one could expect, 
from our $\LL(v)$ matrix in 
the limit $b\rightarrow 0$, as in \eref{eq:b->0}.
The result is a variant of the standard 
$n\times n$
Lax matrix for the periodic Toda lattice \cite{Jimbo85}: 
\beq
\LL(v)=b^{-1}+\LL^{TL}(v)+\Or(b),
\eeq
\bea
\LL^{TL}(v)=&v^{-1}\rme^{q_n-q_1}E_{1n}+vE_{n1} \nonumber \\
  &+\sum_{j=1}^n(p_j+c_j)E_{jj}
   +\sum_{j=1}^{n-1}\rme^{q_j-q_{j+1}}E_{j+1,j}
   +\sum_{j=1}^{n-1}E_{j,j+1}.
\eea

Similarly, from $\MM(v)$ one obtains the corresponding matrix for the Toda
lattice, see \cite{KS5}.

The Poisson brackets for both dual Lax matrices 
$\LL(v)$ can be expressed
in the generalized $r$-matrix form \cite{STS83}
\beq
 \{\one\LL(v_1),\two\LL(v_2)\}
   =[r_{12}(v_1,v_2),\one\LL(v_1)]-[r_{21}(v_1,v_2),\two\LL(v_2)],
\eeq
the `non-unitary' $r$-matrix having the form:
\beq
r_{12}(v_1,v_2)=\frac{1}{v_1-v_2}\,\left(v_2\,\sum_{k\geq j}\,
+\,{v_1}\,\sum_{k<j}\,\right)E_{jk}\otimes E_{kj}
\label{eq:nonun-r}
\eeq
and $r_{21}(v_1,v_2)={\mathcal P}r(v_2,v_1){\mathcal P}$,
where
${\mathcal P}=\sum_{j,k=1}^n E_{jk}\otimes E_{kj}$
is the permutation matrix in $\C^n\otimes\C^n$.

The non-unitary $r$-matrix \eref{eq:nonun-r} in case of Toda's Lax matrix 
can be unitarized by a gauge trans\-for\-ma\-ti\-on:
\beq
\hbox{\frak L}(v)= V\LL^{TL}(v)V^{-1}, \qquad 
V=\sum_{j=1}^n \rme^{q_j/2}E_{jj}
\eeq
getting for the new Lax matrix $\hbox{\frak L}(v)$
the standard unitary $A_{n-1}$-type $r$-matrix
\beq
\hbox{\frak r}_{12}(v_1,v_2)
=\frac{v_1+v_2}{v_1-v_2}\,\sum_{j=1}^n\, E_{jj}\otimes E_{jj}
+\frac{1}{v_1-v_2}\,\left(v_2\,\sum_{k> j}\,
+\,{v_1}\,\sum_{k<j}\,\right)E_{jk}\otimes E_{kj},
\eeq
\beq
\hbox{\frak r}_{12}(v_1,v_2)=-\hbox{\frak r}_{21}(v_1,v_2)
\eeq
\beq
\{\one{\hbox{\frak L}}(v_1),\two{\hbox{\frak L}}(v_2)\}
=[\hbox{\frak r}_{12}(v_1,v_2),
\one{\hbox{\frak L}}(v_1)+\two{\hbox{\frak L}}(v_2)],
\eeq
see, for instance, \cite{Jimbo85}.

\newsection{Quantization}
\setcounter{equation}{0}
\label{sec:quant}

In the quantum case the canonical momenta $X_i$ are replaced with
the differentiations $\dd_i\equiv \dd/\dd x_i$ (having no intent to discuss
the conjugation properties of the observables, we discard the factor 
$\rmi\hbar$ to simplify the notation). To preserve the commutativity of
the Hamiltonians $H_i$ upon quantization one needs to choose the
operator ordering in a special way.

The necessary algebraic framework is given by the Quantum Inverse Scattering
or the $R$-matrix \cite{Bax82,KBI93} method. Defining the local quantum 
Lax matrix
as
\beq
 \ell_i(u)=\left(\begin{array}{cc}
       u-c_i-x_i\dd_i & b x_i \\ -\dd_i & b \end{array}\right)
\label{eq:def-q-ell}
\eeq
one verifies the commutation relation
\beq
R_{12}(u_1-u_2)\one\ell(u_1)\two\ell(u_2)=\two\ell(u_2)\one\ell(u_1)R_{12}(u_1-u_2)
\label{eq:Rll}
\eeq
where
\beq
R_{12}(u)=u+\PP_{12}
\label{eq:def-R}
\eeq
is the standard $SL(2)$-invariant solution to the quantum Yang-Baxter
equation. The quantum Lax operator, or monodromy matrix, $L(u)$ and its
trace $t(u)$ are
defined then by the same formulas \eref{eq:monodromy} and \eref{eq:def-t}
as in the classical case.
{}From \eref{eq:Rll} one then derives in a standard way the similar
commutation relation
\beq
R_{12}(u_1-u_2)\one L(u_1)\two L(u_2)=\two L(u_2)\one L(u_1)R_{12}(u_1-u_2)
\label{eq:RLL}
\eeq
for $L(u)$ from which the commutativity of $t(u)$
\beq
 [t(u_1),t(u_2)]=0
\eeq
follows immediately. The commutative quantum Hamiltonians $H_i$ are defined
then, like in the classical case \eref{eq:def-Hi}, as coefficients of
the polynomial $t(u)$. It is easy to see that $H_i$ is a differential operator
of order $i$ leaving invariant the space 
$\Cx$ of polynomials of $x_1,\ldots x_n$.
In particular, $H_1$ and $H_2$ are given by the formulas \eref{eq:def-H12}
with $N$ and $H$ given by \eref{eq:def-N} and \eref{eq:def-H}, respectively.

The main problem in the quantum case is the spectral problem
for commuting differential operators, quantum Hamiltonians
$\{H_i\}_{i=1}^n$:
\beq
H_i\psi(x_1,\ldots,x_n)=h_i\psi(x_1,\ldots,x_n),\qquad 
\psi(x_1,\ldots,x_n)\in\Cx.
\eeq

One can describe the spectrum and eigenvectors of $H_i$, or, equivalently, 
$t(u)$ using the well-developed machinery of
{\it algebraic Bethe Ansatz} \cite{KBI93}. Defining the vacuum
state $\vac$ as the unit function $\vac(x)\equiv1$
in $\Cx$ we note that
\beq
L_{21}\vac=0, \qquad
L_{11}(u)\vac=\alpha_{11}(u)\vac, \qquad
L_{22}(u)\vac=\alpha_{22}(u)\vac,
\eeq
where
\beq
\alpha_{11}(u)=\prod_{i=1}^n(u-c_i), \qquad
\alpha_{22}(u)=b^n.
\eeq

Defining the {\it Bethe vector} $\psi_{\vec v}(x_1,\ldots,x_n)\in\Cx$
parametrized by $m$ complex numbers $v_j$ as
\beq
\psi_{\vec v}(x_1,\ldots,x_n)\equiv
\ket{v_1,\ldots,v_m}= L_{12}(v_1)\ldots L_{12}(v_m)\vac
\eeq
one can prove \cite{KBI93}, using the commutation relations \eref{eq:RLL}, that
$\ket{v_1,\ldots,v_m}$ is an eigenvector of $t(u)$, for any $u\in\C$, 
if and only if the parameters $v_j$ satisfy the system of algebraic
{\it Bethe equations}
\beq
 \prod_{j=1}^m\frac{v_k-v_j+1}{v_k-v_j-1}
 =-\frac{\alpha_{11}(v_k)}{\alpha_{22}(v_k)}, \qquad
 k=1,\ldots,m
\label{eq:Bethe-eqs}
\eeq
and the corresponding eigenvalue $\tau(u)$ of $t(u)$
\beq
t(u)\ket{v_1,\ldots,v_m}=\tau(u)\ket{v_1,\ldots,v_m}
\eeq
is given by the formula
\beq
\tau(u)=\alpha_{11}(u)\prod_{j=1}^m\frac{u-v_j-1}{u-v_j}
       +\alpha_{22}(u)\prod_{j=1}^m\frac{u-v_j+1}{u-v_j}.
\label{eq:tau}
\eeq

It is usually assumed that Bethe eigenvectors are complete, at least
for generic values of parameters. The proof of the conjecture is, however,
a difficult task, and is available only for a few models, see \cite{KBI93}
for a discussion.

In his seminal study \cite{Bax82}
of the integrable XYZ and XXZ spin chains R J Baxter
has pointed out that the equations similar to ours 
\eref{eq:Bethe-eqs} and \eref{eq:tau} can be reformulated equivalently
as a finite-difference equation in a certain class of holomorphic functions.
Adapting his reasoning to our case we introduce the polynomial
$\phi(\l;\vec v)$ in $\l$ whose zeros are the Bethe parameters $v_j$:
\beq
 \phi(\l;\vec v)=\prod_{j=1}^m(\l-v_j), \qquad \l\in\C.
\label{eq:def-phi}
\eeq

It is easy to see then that the following finite-difference equation
of second order  for $\phi(\l;\vec v)$
\beq
 \phi(\l;\vec v)\tau(\l)=
\alpha_{11}(\l)\phi(\l-1;\vec v)+\alpha_{22}(\l)\phi(\l+1;\vec v)
\label{eq:sep-eq-phi}
\eeq
is equivalent to the system of equations \eref{eq:Bethe-eqs}
for $\{v_j\}_{j=1}^m$ and to the equation \eref{eq:tau}
for $\tau(\l)$. To show this, it is sufficient to divide both
sides of \eref{eq:sep-eq-phi} by $\phi(\l)$ and take residues at
$\l=v_j$. The equation \eref{eq:sep-eq-phi} is called {\it Baxter's} or
{\it separation} equation. The reason for the latter name is that an
identical equation arises when solving the model via the separation of
variables method (see \cite{KS5} for more on relation between 
$Q$-operator and quantum separation of variables).

Now we are able to describe the problem we are going to study in the 
remaining sections of this paper.
We are looking for a one-parameter family of operators $Q_\l$ acting
in $\Cx$ such that $Q_\l$ shares with $t(u)$ the same set of
Bethe eigenvectors, and the eigenvalues $q(\l)$ of $Q_\l$ 
\beq
 Q_\l\ket{v_1,\ldots,v_m}=q(\l)\ket{v_1,\ldots,v_m}
\eeq
are polynomials in $\l$ satisfying Baxter's equation \eref{eq:sep-eq-phi}.
Up to a normalization coefficient $\kappa_{\vec v}$, depending on the
eigenvector, the polynomials $q(\l)$ are proportional to the polynomials
$\phi(\l;\vec v)$ defined by \eref{eq:def-phi}:
\beq
 q(\l)=\kappa_{\vec v}\phi(\l;v_1,\ldots,v_m)
=\kappa_{\vec v}\l^m+\Or(\l^{m-1}), \qquad \l\rightarrow\infty.
\label{eq:def-kappa}
\eeq

Instead of dealing with eigenvectors and eigenvalues
it is more convenient to cha\-rac\-te\-ri\-ze $Q_\l$ by the following
operator identities which are equivalent to the above characterization,
assuming the completeness of Bethe eigenvectors.
We demand that $Q_\l$ commute with $t(u)$ 
\begin{subeqnarray}
    [t(u),Q_\l]&=&0
\label{eq:comm-tQ}
\end{subeqnarray}
and self-commute
\addtocounter{equation}{-1}
\begin{subeqnarray}\addtocounter{subequation}{1}
 [Q_{\l_1},Q_{\l_2}]&=&0,
\label{eq:comm-QQ}
\end{subeqnarray}
as well as satisfy the finite-difference equation
\addtocounter{equation}{-1}
\begin{subeqnarray}\addtocounter{subequation}{2}
\label{eq:Qt}
 Q_\l t(\l)&=&Q_{\l-1}\prod_{i=1}^n(\l-c_i)+b^nQ_{\l+1}.
\end{subeqnarray}

In addition, the eigenvalues of $Q_\l$ should be polynomial in $\l$
\addtocounter{equation}{-1}
\begin{subeqnarray}\addtocounter{subequation}{3}
 &&q(\l)\in\C[\l].
\label{eq:q-poly}
\end{subeqnarray}

The above conditions by no means define $Q_\l$ uniquely. Apparently, one
can construct infinitely many $Q$-operators just by fixing arbitrary
normalization coefficients $\kappa_{\vec v}$ for each eigenvector 
$\ket{\vec v}$ in
\eref{eq:def-kappa}. The difficult problem is to find an explicit
expression for a $Q$-operator. 
Baxter succeeded in solving the problem in case of XYZ
and XXZ spin chains, having given an expression for $Q_\l$ as a trace of
a monodromy matrix \cite{Bax82}. However, his formulas do not survive passing 
to the limiting case of the XXX spin chain, governed by the $SL(2)$
invariant $R$ matrix \eref{eq:def-R}.

In the case of quantum periodic Toda lattice, which is another model
governed by the $R$ matrix \eref{eq:def-R}, a solution was found by Pasquier 
and Gaudin \cite{PG92}. Instead of trying to construct $Q_\l$ as trace of
a monodromy matrix, they considered $Q_\l$ as
an integral operator
\beq
 Q_\l:\psi(\vec x)\mapsto 
 \int\rmd x_1 \ldots\int\rmd x_n\, \QQ_\l(\vec y\mid\vec x)\,\psi(\vec x)
\label{eq:def-kernelQ}
\eeq
having given an explicit expression for its kernel $\QQ_\l(\vec y\mid\vec x)$.
They also discovered an important relation between the kernel 
$\QQ_\l(\vec y\mid\vec x)$
and the generating function $F_\l(\vec y\mid\vec x)$
of the classical \bt\ expressed by the semiclassical formula
\beq
 \QQ_\l(\vec y\mid\vec x)\sim
 \exp\left(-\frac{\rmi}{\hbar}F_\l(\vec y\mid\vec x)\right),
\qquad \hbar\rightarrow0.
\label{eq:Q-semicl}
\eeq

The classical \bt\ $B_\l$ is thus the classical limit of the similarity 
transformation ${\mathcal O}\mapsto Q_\l{\mathcal O}Q_\l^{-1}$.

Recently, it was found \cite{BLZ} how the original Baxter's construction
\cite{Bax82} can be generalized to produce
$Q$-operators for the models governed by the $A_1$-type
$R$ matrices, such as the XXZ spin chain and sine-Gordon model.
According to \cite{BLZ}, $Q_\l$ is constructed as the trace of
a monodromy matrix built of the local Lax operators corresponding, in the
auxiliary space, to the special infinite-dimensional representations
of the quantum group ${\mathcal U}_q[\widehat{sl_2}]$ ({\it $q$-oscillator}
representations).

In the subsequent
sections we construct a $Q$-operator for the quantum DST model
and prove its characteristic properties.
Our approach combines those of \cite{PG92} and of \cite{BLZ}.
Similarly to \cite{BLZ}, we construct our $Q$-operator as the trace of
a monodromy matrix with an infinite-dimensional auxiliary space.
In the spirit of \cite{PG92}, we find $Q_\l$ as an integral operator acting
in $\C[\vec x]$ and present several equivalent expressions for it.

The $Q$-operator being found as an integral operator will give
integral equations for the eigenfunctions $\psi_{\vec v}$.
The advantage of this transformation of the differential 
spectral problem into integral spectral problem is that it gives
an alternative to Bethe representation of multivariable special functions.
The general approach of constructing a $Q$-operator for a given
integrable system will be of even greater importance in situations when
Bethe ansatz does not work. 

\newsection{Construction of the $Q$-operator}
\setcounter{equation}{0}
\label{sec:R-operator}

The structure of $Q_\l$ is similar to that of $t(u)$ given by
\eref{eq:monodromy} and \eref{eq:def-t}. We construct $Q_\l$ as the trace
of a monodromy matrix built of the elementary blocks $\R^{(i)}_{\l-c_i}$.
Suppose that
$\R_\l$ is a linear operator from $\C[s,x]$ into $\C[t,y]$. The spaces
$\C[x]$ and $\C[y]$ are referred to as {\it quantum} spaces and $\C[s]$ and
$\C[t]$, respectively, as {\it auxiliary} ones (see \cite{KBI93}).
To construct $Q_\l$ we introduce $n$
copies $\R_{\l-c_i}^{(i)}$ of $\R_\l$ assuming that
$\R_{\l-c_i}^{(i)}:\C[s_i,x_i]\mapsto\C[s_{i+1},y_i]$
(remember the periodicity convention $n+1\equiv 1$)
and extending $\R_{\l-c_i}^{(i)}$ on $\C[x_j]$
($j\neq i$) as the unit operator. The monodromy matrix
$\R_{\l-c_n}^{(n)}\ldots\R_{\l-c_1}^{(1)}$ acts then from
$\C[s_1,\vec x]$ into $\C[s_1,\vec y]$, and $Q_\l$ is obtained by taking
trace in the auxiliary space $\C[s_1]$:
\beq
 Q_\l=\tr_{s_1}\R_{\l-c_n}^{(n)}\ldots\R_{\l-c_1}^{(1)}.
\label{eq:Q=trRR}
\eeq

Supposing $\R_\l$ to be an integral operator
\beq
 \R_\l:\psi(s,x)\mapsto \int\rmd x\int\rmd s\,
\RR_\l(t,y\mid s,x)\,\psi(s,x)
\label{eq:def-RR}
\eeq
we have for the kernel 
$\QQ_\l(\vec y\mid \vec x)$ of $Q_\l$
\beq
\QQ_\l(\vec y\mid \vec x)=
\int\rmd s_n\ldots\int\rmd s_1
\prod_{i=1}^n \RR_{\l-c_i}(s_{i+1},y_i\mid s_i,x_i).
\label{eq:ker-Q-as-int}
\eeq

To ensure the commutativity $[t(u),Q_\l]=0$ it is sufficient
to demand that $\R_\l$ intertwines
\beq
 \M(u-\l;t,\dd_t)\ell(u;y,\dd_y)\R_\l=
\R_\l\ell(u;x,\dd_x)\M(u-\l;s,\dd_s)
\label{eq:MelR}
\eeq
the local Lax operator $\ell(u)$ and
some other representation $\M(u-\l)$ of the same algebra  \eref{eq:Rll}
\beq
 R(u_1-u_2)\one\M(u_1)\two\M(u_2)=
 \two\M(u_2)\one\M(u_1)R(u_1-u_2)
\label{eq:RMM}
\eeq
with the same $R$ matrix \eref{eq:def-R}. The proof of (\ref{eq:comm-tQ}a)
follows then by a standard argument \cite{FTbook,KBI93}.

Similarly, to prove $[Q_{\l_1},Q_{\l_2}]=0$ (\ref{eq:comm-QQ}b)
it is sufficient to establish
the Yang-Baxter identity
\bea
& \int\rmd t_1 \int\rmd t_2 \int\rmd y\,
 \tilde\RR_{\l_1-\l_2}(w_1,w_2\mid t_1,t_2)
       \RR_{\l_1}(t_1,z\mid s_1,y)
       \RR_{\l_2}(t_2,y\mid s_2,x) \nonumber\\
&= \int\rmd t_1 \int\rmd t_2 \int\rmd y\,
       \RR_{\l_2}(w_2,z\mid t_2,y)
       \RR_{\l_1}(w_1,y\mid t_1,x)
 \tilde\RR_{\l_1-\l_2}(t_1,t_2\mid s_1,s_2)
\label{eq:YBE}
\eea
with some kernel $\tilde\RR_\l$.

The representation $\M(u-\l)$ of the algebra \eref{eq:RMM}
should be chosen in such a way
that the resulting $Q_\l$, as a function of $\l$, 
satisfy Baxter's finite-difference equation 
(\ref{eq:Qt}c)
and have polynomial eigenvalues (\ref{eq:q-poly}d). As we shall show, for this
purpose one can take
\beq
 \M(u;s,\dd_s)=\left(\begin{array}{cc}
      u-s\dd_s & s \\ -\dd_s & 1
   \end{array}\right)
\label{eq:def-qM}
\eeq
coinciding essentially with $\ell(u)$ with $b=1$ and $c_i=0$.
The representation \eref{eq:def-qM} plays for the Yangian ${\mathcal Y}[sl_2]$
the same role as the $q$-oscillator representation plays for the quantum group
${\mathcal U}_q[\widehat{sl_2}]$ in \cite{BLZ}.
Having fixed $\M(u)$
by \eref{eq:def-qM} we get from \eref{eq:MelR} a set of differential
equations for the kernel $\RR_\l(t,y\mid s,x)$ of $\R_\l$
\bea
& \left(\begin{array}{cc}
  u-\l-t\dd_t & t \\ -\dd_t & 1
 \end{array}\right)
 \left(\begin{array}{cc}
  u-y\dd_y & by \\ -\dd_y & b
 \end{array}\right)
\RR_\l(t,y\mid s,x) \nonumber \\
 &=\left(\begin{array}{cc}
  u+1+x\dd_x & bx \\ \dd_x & b
 \end{array}\right)
 \left(\begin{array}{cc}
 u-\l+1+s\dd_s & s \\ \dd_s & 1
 \end{array}\right)
\RR_\l(t,y\mid s,x)
\label{eq:eqs-RR}
\eea
(in the right-hand-side we have used integration by parts and the identities
$\dd_x^*=-\dd_x$, $(x\dd_x)^*=-\dd_x x=-1-x\dd_x$). The equations 
\eref{eq:eqs-RR} determine $\RR_\l$ up to a scalar factor $\rho_\l$:
\beq
 \RR_\l(t,y\mid s,x)=\rho_\l
 \delta(s-by)  y^{-1} \exp\left(\frac{t-x}{y}\right)
\left(\frac{t-x}{y}\right)^{-\l-1}.
\label{eq:RR}
\eeq

It remains to choose the factor $\rho_\l$ in \eref{eq:RR}
and the integration contour in \eref{eq:def-RR} in such a way that
$\R_\l:\C[s,x]\mapsto\C[t,y,\l]$.

We describe first the final formula for $\R_\l$ and equivalent expressions
and then prove the polynomiality property. As the basic definition of $\R_\l$
we choose the following formula:
\beq
\R_\l:\psi(s,x)\mapsto \frac{\rmi}{2\pi}\Gamma(\l+1)\int_\gamma \rmd\xi\,
  \rme^{-\xi}(-\xi)^{-\l-1}\, \psi(by,y\xi+t).
\label{eq:RR-final}
\eeq

\begin{figure}[h]
\begin{picture}(30,10)(-5,-5)
\put(0,0){\circle*{0.4}}
\thicklines
\put(0,0){\line(1,0){25}}
\thinlines
\put(0,0){\makebox(0,-0.3)[rt]{$0$}}
\put(0,0){\oval(6,6)[l]}
\put(0,3){\line(1,0){20}}
\put(0,-3){\line(1,0){20}}
\put(15,3){\vector(-1,0){5}}
\put(5,-3){\vector(1,0){5}}
\put(11,3.6){\makebox(0,0)[b]{$\gamma$}}
\end{picture}

\caption[tri]{Integration contour $\gamma$}
\label{fig: contour}
\end{figure}
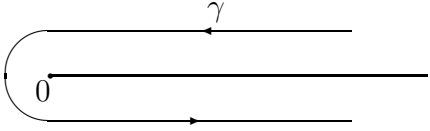

The infinite integration contour $\gamma$ is shown on Figure
\ref{fig: contour}.
The branch of the many-valued function
$(-\xi)^{-\l-1}$ in \eref{eq:RR-final}
is fixed by making a cut along $(0,\infty)$ and assuming
that $-\pi\leq\arg(-\xi)\leq\pi$.

{}From \eref{eq:RR-final} it is apparent that $\R_\l$, as a function of $\l$,
is analytic in $\C$ except the poles $\l=-1,-2,\ldots$ of the factor
$\Gamma(\l+1)$. As shown below, in fact $\R_\l$ continues analytically on
the whole complex plane.

Indeed, for $\re\l<0$ one can pull the contour
$\gamma$ over the cut $(0,\infty)$ and replace $\int_\gamma\rmd\xi\,f(\xi)$
with $\int_0^\infty\rmd\xi\,[f(\xi-\rmi0)-f(\xi+\rmi0)]$ which results in the
formula:
\beq
  \R_\l:\psi(s,x)\mapsto \frac{1}{\Gamma(-\l)}\int_0^\infty\rmd\xi\,
  \rme^{-\xi}\xi^{-\l-1}\, \psi(by,y\xi+t), \qquad \re\l<0
\label{eq:RR-0inf}
\eeq
which is analytic in $\l=-1,-2,\ldots$. The branch of $\xi^{-\l-1}$ 
in \eref{eq:RR-0inf} is fixed by the condition $\arg\,\xi=0$.

To put $\R_\l$ in the form \eref{eq:def-RR} convenient for checking the
intertwining relation \eref{eq:eqs-RR} one has to make the change of
variables $x=y\xi+t$ in \eref{eq:RR-final}. The result is given by the
formula \eref{eq:def-RR} with the kernel $\RR_\l$ given by the expression
\eref{eq:RR} with the scalar factor
$\rho_\l=\frac{\rmi}{2\pi}\Gamma(\l+1)$
and integration in $x$ taken over the contour
$\gamma^\prime=y\gamma+t$. As for the integration contour in $s$, it needs
only to pass through the point $s=by$ because of the factor $\delta(s-by)$ in
$\RR_\l$.

In the same way, from \eref{eq:RR-0inf} one obtains again the
formula \eref{eq:def-RR} with the kernel
\eref{eq:RR} with the different scalar factor
$\rho_\l=1/\Gamma(-\l)$
and integration in $x$ taken over the ray starting from $x=t$ and going in
the direction of $y/\left|y\right|$.

Now we have the full description of the operator $\R_\l$ and can start
to study its properties. By construction, $\R_\l$ satisfies the relation
\eref{eq:MelR} from which the commutativity $[t(u),Q_\l]=0$
(\ref{eq:comm-tQ}a) follows.
By direct calculation one can establish also the Yang-Baxter identity
\eref{eq:YBE} with the kernel $\tilde\RR_\l\equiv\left.\RR_\l\right|_{b=1}$
proving thus the commutativity $[Q_{\l_1},Q_{\l_2}]=0$ (\ref{eq:comm-QQ}b).
The proof of the remaining properties of $Q_\l$ from the list presented
in section \ref{sec:quant} is given in sections \ref{sec:Q polynom}
and \ref{sec:Q Baxter}.

We conclude this section with giving an alternative description of $\R_\l$
in terms of the polynomial bases which complements the above ones in terms of
integral operators.

To calculate explicitely the action of $\R_\l$ on the monomial basis
$s^kx^j$ in $\C[s,x]$ one puts $\psi(s,x)=s^kx^j$ in \eref{eq:RR-final},
then expands the binomial $(y\xi+t)^j$ and
applies, termwise, Hankel's integral formula \cite{BE}
\beq
  \int_\gamma\rmd\xi\, \rme^{-\xi}(-\xi)^{\nu-1}=
 -\frac{2\pi\rmi}{\Gamma(1-\nu)}.
\label{eq:Hankel}
\eeq

Using the Pochhammer symbol
$(c)_m\equiv \Gamma(c+m)/\Gamma(c)=c(c+1)\ldots(c+m-1)$
one can write down the result as
\beq
 \R_\l:s^kx^j\mapsto
 \sum_{m=0}^j \binom{j}{m}(-\l)_m t^{j-m}y^{m+k}b^k
 =t^j b^k C_j(\l;t/y)
\label{eq:Rsk-Charlier}
\eeq
where $C_m(\l;b)$ are the so-called {\it Charlier polynomials}
\cite{BE,Koe}
$$
C_m(\l;b)={}_2F_0\left[\matrix{-m, \; -\lambda \cr -}; -b^{-1}\right].
$$

The formula \eref{eq:Rsk-Charlier} proves the polynomiality
$\R_\l:\C[s,x]\mapsto\C[t,y,\l]$.
Note that the normalization of $\R_\l$ is chosen in such a way that
$\R_\l:1\mapsto1$.

The action of $\R_\l$ on polynomials can be described in an even more
compact way. Substituting $\psi(s,x)=s^k(x-t)^j$ into \eref{eq:RR-final}
and using again Hankel's formula \eref{eq:Hankel} one obtains the most
economic description of $\R_\l$
\beq
 \R_\l:s^k(x-t)^j\mapsto y^{j+k}(-\l)_j b^k.
\label{eq:R-poly}
\eeq

In the end of the next section we will discuss the formula \eref{eq:R-poly}
and similar ones in more detail.

\newsection{Analytical properties of the $Q$-operator}
\setcounter{equation}{0}
\label{sec:Q polynom}

To produce a description of $Q_\l$ as an integral operator
\eref{eq:def-kernelQ} we substitute
the expression \eref{eq:RR} for the kernel $\RR_\l$ found in the previous
section into the formula \eref{eq:ker-Q-as-int}. The integration in
$s_i$ is easily performed due to the delta-function factors in $\RR_\l$
and, corresponding to the two choices of the factor $\rho_\l$ in \eref{eq:RR}
and integration contour in \eref{eq:def-RR}, we obtain two 
equivalent descriptions of $Q_\l$.

The first formula for $Q_\l$ is given by \eref{eq:def-kernelQ} with the
kernel $\QQ_\l(\vec y\mid\vec x)$
\beq
 \QQ_\l(\vec y\mid\vec x)=\prod_{i=1}^n
 w_i(\l;y_{i+1},y_i,x_i)
\label{eq:def-QQ}
\eeq
where
\beq
 w_i(\l;y_{i+1},y_i,x_i)=
 \frac{\rmi}{2\pi}\Gamma(\l+1-c_i)\,
 y_i^{-1}\left(\frac{by_{i+1}-x_i}{y_i}\right)^{c_i-\l-1}
 \exp\left(\frac{by_{i+1}-x_i}{y_i}\right)
\label{eq:def-w}
\eeq
and integration in $x_i$ is taken over the contour
$\gamma_i=y_i\gamma+by_{i+1}$
whereas the contour $\gamma$ is defined in the previous section.

The alternative formula is given again by \eref{eq:def-kernelQ} with the
kernel $\tilde\QQ_\l(\vec y\mid\vec x)$
\beq
 \tilde\QQ_\l(\vec y\mid\vec x)=\prod_{i=1}^n
 \tilde w_i(\l;y_{i+1},y_i,x_i)
\label{eq:def-tQQ}
\eeq
where
\beq
 \tilde w_i(\l;y_{i+1},y_i,x_i)=
 \frac{y^{-1}_i}{\Gamma(c_i-\l)}
 \left(\frac{x_i-by_{i+1}}{y_i}\right)^{c_i-\l-1}
 \exp\left(\frac{by_{i+1}-x_i}{y_i}\right)
\label{eq:def-wt}
\eeq
and integration in $x_i$ is taken over the straight ray starting from
$x_i=by_{i+1}$ and extending to infinity in the direction
$y_i/\left|y_i\right|$.

Note that the kernels $\QQ_\l(\vec y\mid\vec x)$
and $\tilde\QQ_\l(\vec y\mid\vec x)$ satisfy the semiclassical condition
\eref{eq:Q-semicl} which, taking into account our quantization convention
$-\rmi\hbar=1$, takes the fol\-low\-ing form (up to insignificant
$\l$-depending factors):
\beq
\QQ_\l(\vec y\mid\vec x)\simeq\exp\bigl(-F_\l(\vec y\mid\vec x)\bigr)
\label{eq:semicl-Q}
\eeq
with the generating function of \bt\ given by \eref{eq:def-F}.
Actually, the semiclassical approximation is almost exact, up to a minor
quantum correction $c_i-\l\mapsto c_i-\l-1$. This fact supports our
thesis on the intermediate place of the DST model, with regard to complexity,
between the Toda lattice and the generic XXX  spin chain.
For comparison, in the case
of  the Toda lattice the semiclassical formula for
$\QQ_\l(\vec y\mid\vec x)$ is plainly exact \cite{PG92}, whereas for the XXX
spin chain there is little hope for such a simple result.

For the purposes of the present section we shall need the expressions for
$Q_\l$ similar to the formulas \eref{eq:RR-final} and \eref{eq:RR-0inf}
for $\R_\l$. The corresponding formulas are produced, respectively,
from \eref{eq:def-QQ} and \eref{eq:def-tQQ} by the change of variables
$x_i=y_i\xi_i+by_{i+1}$.

The analog of \eref{eq:RR-final} is the formula
\beq
 Q_\l:\psi(\vec x)\mapsto
 \int_\gamma\rmd\xi_1\ldots \int_\gamma\rmd\xi_n\,
 \WW_\l(\vec\xi)\,\psi(\ldots,y_i\xi_i+by_{i+1},\ldots),
\label{eq:Qpsi-W}
\eeq
where
\beq
 \WW_\l(\vec\xi)=\prod_{i=1}^n \frac{\rmi}{2\pi}\Gamma(\l+1-c_i)
 \rme^{-\xi_i}(-\xi_i)^{c_i-\l-1}.
\label{eq:def-W}
\eeq
valid for any complex $\l$, except the poles
$\l=c_i-k$, ($i=1,\ldots,n$; $k=1,2,\ldots$) of $\Gamma(\l+1-c_i)$.
The branch of each of many-valued functions
$(-\xi_i)^{c_i-\l-1}$ in \eref{eq:Qpsi-W}
is fixed by making a cut along $(0,\infty)$ and assuming
that $-\pi\leq\arg(-\xi_i)\leq\pi$.

The analog of \eref{eq:RR-0inf} is the formula
\beq
 Q_\l:\psi(\vec x)\mapsto
 \int_0^\infty\rmd\xi_1\ldots \int_0^\infty\rmd\xi_n\,
 \tilde\WW_\l(\vec\xi)\,\psi(\ldots,y_i\xi_i+by_{i+1},\ldots),
\label{eq:Qpsi-Wt}
\eeq
with the kernel $\tilde\WW_\l$
\beq
 \tilde\WW_\l(\vec\xi)=\prod_{i=1}^n
 \frac{\rme^{-\xi_i}\,\xi_i^{c_i-\l-1}}{\Gamma(c_i-\l)}
\label{eq:def-Wt}
\eeq
valid for $\re\l<\min\re c_i$. Together, the formulas \eref{eq:Qpsi-W}
and \eref{eq:Qpsi-Wt} define $Q_\l$ as a holomorphic function of $\l\in\C$.

In the rest of this section we will show,
that $Q_\l$ maps polynomials in $x$ into polynomials in $y$ and $\l$
and derive explicit formulas for its action on the monomial basis in $\Cx$.

Before considering the general case
we will give a brief account of the simplest $n=1$ case. 
In this case
we have only one variable $x\equiv x_1$, the Lax matrix
simplifies to $L(u)=\ell(u)$, so, without loss of generality, one can put
$c_1=0$. Trace of $L(u)$ gives rise to only one integral of motion
(number of particles $N$)
\beq
 t(u)\equiv\tr L(u)=u-N+b, \qquad N=x\dd.
\label{eq:defN1}
\eeq

We assume that $N$ acts in the space $\C[x]$ of polynomials of $x$ spanned by
the eigenbasis $\{x^m\}_{m=0}^\infty$ of $N$
\beq
N:x^m\mapsto mx^m,\qquad m=0,1,2,\ldots.
\eeq

For $n=1$ and $c=0$ the formula \eref{eq:Qpsi-W} defining the $Q$ operator
turns into
\beq
 Q_\l:\psi(x)\mapsto 
 \frac{\rmi}{2\pi}\Gamma(\l+1)
 \int_\gamma\rmd\xi\,
 \rme^{-\xi}(-\xi)^{-\l-1}\,\psi(y(\xi+b)), \qquad
 \l\neq -1,-2,\ldots
\label{eq:Qpsi-W1}
\eeq
and \eref{eq:Qpsi-Wt}, respectively, into
\beq
 Q_\l:\psi(x)\mapsto 
\frac{1}{\Gamma(-\l)}\int_0^\infty\rmd\xi\,
\rme^{-\xi}\xi^{-\l-1} \,\psi(y(\xi+b)), \qquad
 \re\l<0.
\label{eq:Qpsi-Wt1}
\eeq

Similarly, from \eref{eq:def-QQ} and \eref{eq:def-tQQ} one gets, respectively,
\beq
  Q_\l:\psi(x)\mapsto 
  \frac{\rmi\rme^b}{2\pi}\Gamma(\l+1)
\int_{\gamma^\prime}\rmd x\,
 y^{-1}\left(b-\frac{x}{y}\right)^{-\l-1}\rme^{-x/y}\,\psi(x), \qquad
 \gamma^\prime=y(\gamma+b)
\eeq
and
\beq
  Q_\l:\psi(x)\mapsto 
 \frac{\rme^b}{\Gamma(-\l)}\int_{by}^\infty\rmd x\,
  y^{-1}\left(\frac{x}{y}-b\right)^{-\l-1}\rme^{-x/y}\,\psi(x), \qquad
 y>0.
\eeq

To calculate explicitly the action of $Q_\l$ on the basis $x^m$ one puts
$\psi(x)=x^m$ in \eref{eq:Qpsi-W1}, then expands the binomial $(\xi+b)^m$
and applies, termwise, Hankel's integral formula \eref{eq:Hankel}.
This calculation is very similar to the calculation of $\R_\l s^kx^j$
given by the formula \eref{eq:Rsk-Charlier}.
The result is that the monomials
$\{x^m\}_{m=0}^\infty$ are the eigenvectors of $Q_\l$:
\beq
 Q_\l:x^m\mapsto q_m(\l)y^m,
\label{eq:Qq}
\eeq
the corresponding eigenvalues $q_m(\l)$ being
polynomials in $\l$ of degree $m$ expressed in terms of the Charlier
polynomials $C_m(\l;b)$ as
\beq
  q_m(\l)= \sum_{j=0}^m \binom{m}{j}(-\l)_j b^{m-j}=b^m C_m(\l;b)
\label{eq:expr-qm}
\eeq
(compare to \eref{eq:Rsk-Charlier}).

As an im\-me\-di\-ate consequence,
we have the commutativity $[Q_\l,N]=0$, as well as
(\ref{eq:comm-tQ}a) and (\ref{eq:comm-QQ}b).
Another corollary is that $Q_\l$ maps $\C[x]$ into $\C[y,\l]$.
Note that formula \eref{eq:expr-qm} implies the normalization
$Q_\l:1\mapsto1$.

One can use the integral operator $Q_\l$ to derive few
well known formulas for the orthogonal Charlier polynomials.
For instance, putting $\psi(x)=x^m$ and $y=1$ in \eref{eq:Qpsi-W1} or in
\eref{eq:Qpsi-Wt1} one obtains integral representations for Charlier 
polynomials:
\beq
   C_m(\l;b)= \frac{\rmi}{2\pi}\Gamma(\l+1)\int_\gamma \rmd\xi\,
 \rme^{_\xi}(-\xi)^{-\l-1}\left(1+\frac{\xi}{b}\right)^m,
\eeq
and, respectively,
\beq
 C_m(\l;b)=\frac{1}{\Gamma(-\l)}\int_0^\infty \rmd\xi\,
   \rme^{-\xi} \xi^{-\l-1}\left(1+\frac{\xi}{b}\right)^m
\eeq
(see \cite{BE}).

Equating $\psi(x)$ in \eref{eq:Qpsi-W1} or \eref{eq:Qpsi-Wt1}
with the generating function 
$\rme^{tx}=\sum_{m=0}^\infty x^mt^m/m!$ of the monomials $x^m$ and
taking the integral
one gets the generating function of Charlier polynomials
\beq
 \rme^t\left(1-\frac{t}{b}\right)^\l=
 \sum_{m=0}^\infty \frac{t^m}{m!} C_m(\l;b).
\label{eq:genf-Ch}
\eeq

The recurrence relation for the Charlier
polynomials \cite{Koe,BE} is equivalent to the finite-difference equation
for the polynomials $q_i(\l)$
\beq
 (\l-i+b)q_i(\l)=bq_i(\l+1)+\l q_i(\l-1), 
\label{eq:sepeq1}
\eeq
which coincides with Baxter's equation \eref{eq:sep-eq-phi} for $n=1$
and proves, for $n=1$, the operator relation
(\ref{eq:Qt}c).

{}From the explicit expression \eref{eq:expr-qm} for the polynomials $q_m(\l)$
we conclude that they are normalized by the condition $q_m(0)=b^m$, or,
alternatively, $q_m(\l)=(-\l)^m+\Or(\l^{m-1})$, as $\l\rightarrow\infty$.
In terms of the operator $Q_\l$ it is equivalent to
\beq
 Q_0=b^N
\label{eq:Q0}
\eeq
(see \eref{eq:defN1} for the definition of $N$)
and, respectively, to
\beq
 Q_\l=(-\l)^N+\Or(\l^{N-1}).
\label{eq:Q-asympt}
\eeq

The generalization of the above results to the multivariable case
is quite straight\-forward. To calculate explicitly the action of $Q_\l$
on the monomial basis $x_1^{m_1}\ldots x_n^{m_n}$ in $\Cx$ one
substitutes $\psi(\vec x)=x_1^{m_1}\ldots x_n^{m_n}$ into \eref{eq:Qpsi-W},
then expands
the bi\-no\-mi\-als $(y_i\xi_i+by_{i+1})^{m_i}$ and uses termwise Hankel's
integral formula \eref{eq:Hankel}. Recalling the de\-fi\-ni\-tion 
\eref{eq:expr-qm}
of Charlier polynomials, one obtains the following expression:
\beq
 Q_\l:x_1^{m_1}\ldots x_n^{m_n}\mapsto
 \prod_{i=1}^n b^{m_i} y_{i+1}^{m_i} C_{m_i}(\l-c_i;by_{i+1}/y_i),
\label{eq:actQ-xn}
\eeq
from which it follows immediately that the normalization
condition $Q_\l:1\mapsto1$ holds and that $Q_\l$ maps $\Cx$ into
$\C[\vec y,\l]$. The polynomiality of matrix elements of $Q_\l$ combined
with the commutativity $[Q_{\l_1},Q_{\l_2}]$ (\ref{eq:comm-QQ}b) proves
the polynomiality (\ref{eq:q-poly}d) of the eigenvalues of $Q_\l$.

The formula \eref{eq:actQ-xn} also allows to 
determine the normalization \eref{eq:def-kappa} 
of the eigenvalues of $Q_\l$.
Taking the limit $\l\rightarrow\infty$ in \eref{eq:actQ-xn} and using the
asymptotics $C_m(\l;b)=(-\l/b)^m+\Or(\l^{m-1})$
we conclude that, as in the $n=1$ case,
$Q_\l$ has the asymptotics \eref{eq:Q-asympt} with the operator $N$ given by 
\eref{eq:def-N}.
In contrast, the equality
\eref{eq:Q0}, generally speaking, cannot be generalized to $n>1$, with the
exception of the homogeneous chain case $c_i\equiv0$, $i=1,\ldots,n$, 
when it is replaced by
\beq
 Q_0=b^N U
\eeq
where $U$ is the translation operator $U:x_i\rightarrow y_{i+1}$.

As a final remark of this section, we point out yet another way
of writing down the action of $Q_\l$.
Substituting $\psi(\vec x)$ in \eref{eq:Qpsi-W} with the polynomials
$\om_{\vec y}^{\vec m}\in\Cx$
\be
 \om_{\vec y}^{\vec m}(\vec x)=\prod_{i=1}^n (x_i-by_{i+1})^{m_i}
\ee
parametrized by the multi-index $\vec m=(m_1,\ldots,m_n)$ and a vector
$\vec y=(y_1,\ldots,y_n)$ we obtain, after performing the integrations,
an elegant formula for the action of $Q_\l$ on $\om_{\vec y}^{\vec m}$:
\beq
 Q_\l:\prod_{i=1}^n(x_i-by_{i+1})^{m_i}\mapsto
      \prod_{i=1}^n(c_i-\l)_{m_i}y_i^{m_i}.
\label{eq:actQ-chi-n}
\eeq

The formula \eref{eq:actQ-chi-n} seems to provide the most compact way
to encode the action of $Q_\l$  on polynomials (compare to the formula
\eref{eq:R-poly} for the action of $\R_\l$).
Some caution is necessary,
however, when using it, since the parameters $\vec y$ in
$\om_{\vec y}^{\vec m}$ coincide with the variables in the target space
$\Cy$ of $Q_\l$. One way of interpreting \eref{eq:actQ-chi-n} is
to consider its left-hand-side as a shorthand notation for
$[Q_\l\om_{\vec z}^{\vec m}]_{\vec z=\vec y}$. Another possibility is
to extend the operator $Q_\l$ onto the polynomial ring $\C[\vec x,\vec y]$
assuming that it acts on $\vec y$ trivially:
$Q_\l(\psi(x)\varphi(y))=\varphi(y)Q_\l(\psi(x))$.
Formulas similar to \eref{eq:actQ-chi-n} also arise in the separation
of variables for Macdonald polynomials \cite{KS2}.

It is a challenging problem to take the formulas \eref{eq:R-poly}  and
\eref{eq:actQ-chi-n}
as definitions of $\R_\l$ and $Q_\l$, respectively,
and to build the theory of $Q_\l$ in an entirely algebraic way.

\newsection{Baxter's equation}
\setcounter{equation}{0}
\label{sec:Q Baxter}

In the previous sections we have proved all the properties of $Q_\l$ from
the list given in section \ref{sec:quant} except Baxter's
difference equation (\ref{eq:Qt}c). 
In this section we give a proof of the 
identity (\ref{eq:Qt}c) based on the ideas of \cite{PG92}.

The best suited for our purposes realization of $Q_\l$ is that given by
the formulas \eref{eq:def-kernelQ} and \eref{eq:def-QQ}.
Recalling that
$t(u)=\tr L(u)$ and that $L(u)$ is a $2\times2$ matrix whose entries are
differential operators in $x_i$, we can transform the left-hand-side of
(\ref{eq:Qt}c) as follows
\be
 [Q_\l t(\l)\psi](\vec y)
 = \tr [Q_\l L(\l)\psi](\vec y)
 = \tr \int\rmd x^n\, \QQ_\l(\vec y\mid\vec x)L(\l)\psi(\vec x).
\ee

Performing integration by parts we obtain
\beq
 [Q_\l t(\l)\psi](\vec y)
 = \tr \int\rmd x^n\, [L^*(\l)\QQ_\l(\vec y\mid\vec x)]\psi(\vec x)
\eeq
where $L^*(\l)$  is the matrix composed of adjoint differential operators
$(L_{jk})^*=L_{jk}^*$. For example, $\dd^*=-\dd$,  $(x\dd)^*=-\dd x=-x\dd-1$,
and so on.

Using the factorization \eref{eq:monodromy} of $L(\l)$ into the product of
elementary Lax matrices $\ell_i(\l)$
and the factorization \eref{eq:def-QQ} of the kernel
$\QQ_\l(\vec y\mid\vec x)$ into the factors $w_i$ \eref{eq:def-w}
we can represent the  kernel of the
integral operator $Q_\l t(\l)$ as
\beq
 \bigl[Q_\l  t(\l)\bigr](\vec y\mid\vec x)=
    \tr\ell_n^*(\l)\ldots\ell_1^*(\l)\prod_{i=1}^n w_i
=\tr \Bigl(\ell_n^*(\l)w_{n}\Bigr)\ldots\Bigl(\ell_1^*(\l)w_{1}\Bigr),
\label{eq:factor-Qt}
\eeq
where
\beq
 \ell_i^*(\l)=\left(\begin{array}{cc}
        \l-c_i+1+x_i\dd_{x_i} & bx_i \\
        \dd_{x_i} & b
      \end{array}\right).
\eeq

The possibility of the factorization
\eref{eq:factor-Qt} of $\bigl[Q_\l L(\l)\bigr](\vec y\mid\vec x)$
depends crucially on the fact that the factors $w_i$ \eref{eq:def-w}
depend each only on one variable $x_i$. That is why we take the
left-hand-side of (\ref{eq:Qt}c) to be $Q_\l t(\l)$ rather than
$t(\l)Q_\l$.

The rest of the proof parallels that of the spectrality property for the
classical case given in section \ref{sec:classic}.
Introducing matrices $\tilde\ell_i(\l)$ by the equality
$\ell_i^*(\l)w_i  =w_i\tilde\ell_i(\l)$ and noting that
\beq
 \dd_{x_i}\ln w_i(y_{i+1},y_i,x_i)=
    \frac{c_i-\l-1}{x_i-by_{i+1}}-\frac{1}{y_i}.
\label{eq:def-dw}
\eeq
we obtain
\beq
\tilde\ell_i(\l)=
  \left(\begin{array}{cc}
     \l-c_i+1+x_i\dd_{x_i}\ln w_i & bx_i \\
    \dd_{x_i}\ln w_i & b
  \end{array}\right)
 =\left(\begin{array}{cc} 
\dfrac{b(c_i-\l-1)y_{i+1}}{x_i-by_{i+1}}-\dfrac{x_i}{y_i} & bx_i \\
     \dfrac{c_i-\l-1}{x_i-by_{i+1}}-\dfrac{1}{y_{i}} & b \end{array}\right)
\label{eq:def-tl}
\eeq
and
\beq\
 \bigl[Q_\l  t(\l)\bigr](\vec y\mid\vec x)= \QQ_\l(\vec y\mid\vec x)\,
    \tr\tilde\ell_n(\l)\ldots\tilde\ell_1(\l)
  \equiv\QQ_\l(\vec y\mid\vec x)\,\tr\tilde L(\l).
\label{eq:tltl}
\eeq

We note then that the gauge transformation 
$\tilde\ell_i(\l)\mapsto N_{i+1}^{-1}\tilde\ell_i(\l)N_i$ with the gauge
matrix 
\beq
 N_i=\left(\begin{array}{cc} 1 & by_i \\ 0 & 1 \end{array}\right)
\eeq
leaves $\tr\tilde L(\l)$ invariant while making $\tilde\ell_i(\l)$ and,
consequently, $\tilde L(\l)$ triangular:
\bea
 N_{i+1}^{-1}\tilde\ell_i(\l)N_i&=&
 \left(\begin{array}{cc} -\dfrac{x_i-by_{i+1}}{y_i} & 0 \\
     \dfrac{c_i-\l-1}{x_i-by_{i+1}}-\dfrac{1}{y_{i}} &
    \dfrac{b(c_i-\l-1)y_i}{x_i-by_{i+1}} \end{array}\right) \nonumber \\
 &=&\left(\begin{array}{cc} (\l-c_i)\dfrac{w_i(\l-1)}{w_i(\l)} & 0 \\
        \dfrac{c_i-\l-1}{x_i-by_{i+1}}-\dfrac{1}{y_{i}} &   
     \dfrac{bw_i(\l+1)}{w_i(\l)} \end{array}\right),
\eea
where we used the identities
\beq
 \frac{w_i(\l+1)}{w_i(\l)}=\frac{(c_i-\l-1)y_i}{x_i-by_{i+1}}, \qquad
 \frac{w_i(\l-1)}{w_i(\l)}=\frac{x_i-by_{i+1}}{(c_i-\l)y_i}.
\eeq

As a result, we  get the equality
\beq
  \tr \tilde L(\l)=
 b^n \prod_{i=1}^n \frac{w_i(\l+1)}{w_i(\l)}
+\prod_{i=1}^n (\l-c_i)\frac{w_i(\l-1)}{w_i(\l)},
\label{eq:tr-tell}
\eeq
which, obviously, proves  (\ref{eq:Qt}c).

\newsection{Discussion}
\setcounter{equation}{0}
\label{sec:disc}

On the example of the quantum integrable DST model we have shown that
the construction of $Q$-operator as an integral operator, in the style of
the paper \cite{PG92}, and as the trace of a monodromy matrix with a special
representation of the quantum group corresponding to the auxiliary space,
in the style of papers \cite{BLZ}, can be combined naturally within
an unified approach. The same approach can be applied to other integrable
models which are more general than DST model, 
such as the generic XXX magnetic chain. This work is in progress and
the results will be reported in a separate paper. For a particular case
of the homogeneous XXX chain a $Q$-operator was recently constructed in
\cite{Der99}.

Another interesting problem is to build the theory of $Q$-operator
in a purely algebraic manner starting from the formulas \eref{eq:R-poly}
and \eref{eq:actQ-chi-n}.

In \cite{BLZ} it is argued that for the models governed by
the $A_1$ type $R$ matrices there exist two different $Q$-operators
corresponding to two different $q$-oscilator representations of
${\mathcal U}_q[\widehat{sl_2}]$. Their eigenvalues correspond, respectively,
to two linearly
independent solutions of Baxter's difference equations analogous to
\eref{eq:sep-eq-phi}.
In the case of DST model the second $Q$-operator can be obtained if we
choose in the formula \eref{eq:MelR} another representation
$\M(u-\l)$ of the algebra  \eref{eq:RMM}, namely
$\tilde\M(u-\l)\sim-\M^{-1}(\l-u)$
\beq
 \tilde\M(u;s,\dd_s)=\left(\begin{array}{cc}
    -1 & s \\ -\dd_s & u+s\dd_s
   \end{array}\right).
\eeq

The corresponding $Q$-operator has, however, more complex nature 
than the one studied in the present paper. 
Its eigenvalues, for example, are not polynomial in $\l$. 
The problem is currently under study.

We can point out the following application of our 
results to the theory of special functions of many
variables. Notice that the eigenfunctions of the 
quantum DST Hamiltonians are multivariable polynomials.
The obtained family of integral equations for those polynomials 
provided by the $Q_\l$-operator supplements their representation
as Bethe vectors and can be used in efficient numerical 
calculations of these special functions, for instance,
solving integral equations by iterations. Simple
considerations of the $n=2$ case show that we deal 
with the discrete multivariable analogues of the 
Heun polynomials. For special functions of such complexity
the found integral equations might be the only explicit
representations to exist because there is no hope to
get, for instance, an integral representation.
So, the found integral equations for the special functions
which were initially defined as eigenfunctions of the commuting
differential operators can be used first, as already
remarked, for generating advanced numerical methods 
of their calculation, and, secondly, for finding
various asymptotics. These applications are being worked
on.

\section*{Acknowledgements}

The authors wish to acknowledge the support of EPSRC and INTAS.



\begin{thebibliography}{12}
\bibitem{ELS85} Eilbeck J C, Lomdahl P S, and A~C Scott A C 1985 
The discrete self-trapping equation
{\it Physica D} {\bf 16} 318--338

\bibitem{chaos1} De Filippo S, Fusco Girard M and Salerno M 1989
{\it Nonlinearity} {\bf 2} 477

\bibitem{chaos2} Cruzeiro-Hansson L, Feddersen H, Flesch R, Christiansen P L, 
Salerno M and Scott A C 1990 
Classical and quantum analysis of chaos in the discrete self-trapping equation
{\it Phys.\ Rev.\ B} {\bf 42} 522--526

\bibitem{davydov} Davydov A S and Kislukha N I 1973
{\it Phys.\ Status Solidi B} {\bf 59} 465

\bibitem{fs90} Finlayson N and Stegeman G I 1990 
Spatial switching, instabilities, and chaos in a 3-wave-guide nonlinear
directional coupler
{\it Appl.\ Phys.\ Lett.} {\bf 56} 2276--2278

\bibitem{kc86} Kenkre V M and Campbell D K 1986 
Self-trapping on a dimer --- time-dependent solutions of a discrete nonlinear
Schr\"{o}dinger equation
{\it Phys.\ Rev.\ B} {\bf 34} 4595--4961

\bibitem{se86} Scott A C and Eilbeck J C 1986 
The quantized discrete self-trapping equation
{\it Phys.\ Lett.\ A}, {\bf 119} 60--64 

\bibitem{esks} Enolskii V Z, Salerno M, Kostov N A and Scott A C 1991 
Alternative quantizations of the discrete self-trapping dimer
{\it Physica Scripta} {\bf 43} 229--235

\bibitem{EKS93} Enol'skii V Z, Kuznetsov V B and Salerno M 1993
On the quantum inverse scattering method for the DST dimer
{\it Physica D} {\bf 68} 138--152


\bibitem{CJK93} Christiansen P L, J{\o}rgensen M F and Kuznetsov V B 1993
On integrable systems close to the Toda lattice
{\it Lett.\ Math.\ Phys.\ }{\bf 29} 165--173

\bibitem{Bax82} Baxter R I 1982 {\it Exactly Solved Models in Statistical
Mechanics} (London: Academic) ch 9--10

\bibitem{PG92} Pasquier V and Gaudin M 1992 The periodic Toda chain
and a matrix generalization of the Bessel function recursion relations
{\it J.\ Phys.\ A: Math.\ Gen.\ }{\bf 25} 5243--5252

\bibitem{KS5} Kuznetsov V B and Sklyanin E K 1998
On B\"{a}cklund transformations for many-body systems
{\it J.\ Phys.\ A: Math.\ Gen.\ }{\bf 31} 2241--2251

\bibitem{BLZ} Bazhanov V, Lukyanov S and Zamolodchikov A 1997
Integrable Structure of Conformal Field Theory II. Q-operator and DDV equation
{\it Commun.\ Math.\ Phys.\ }{\bf 190} 247--278
\item[]
Antonov A and Feigin B 1997
Quantum Group Representations and Baxter Equation
{\it Phys.\ Lett.\ }{\bf B392} 115-122

\bibitem{FTbook} Faddeev L D and Takhtajan L A 1987
{\it Hamiltonian Methods in the Theory of solitons} (Berlin: Springer-Verlag)

\bibitem{Skl51} Sklyanin E K 1999a Canonicity of \bt: $r$-matrix approach. I,
preprint LPENSL-Th 05/99; {\sf solv-int/9903016}.

\bibitem{Skl52} Sklyanin E K 1999b Canonicity of \bt: $r$-matrix approach. 
II, preprint LPENSL-Th 06/99; {\sf solv-int/9903017}; to be published in:
{\sl Trudy MIAN, v. 226 (1999), Moscow, Nauka.}

\bibitem{dual} Adams M R, Harnad J and Hurtubise J 1990 Dual moment maps to
loop algebras {\it\LMP} {\bf 20} 294--308

\bibitem{Jimbo85} Jimbo M 1985 
Quantum R matrix for the generalized Toda system
{\it Commun.\ Math.\ Phys.\ } {\bf 102} 527--547

\bibitem{STS83} Semenov-Tian-Shansky M A 1983 What is classical $r$-matrix?
{\it Funct.\ Anal.\ Appl.\ } {\bf 17} 259--272

\bibitem{KBI93} Korepin V A, Bogoliubov N M and Izergin A G 1993
{\it Quantum Inverse Scattering Method and Correlation Functions} 
(Cambridge: Cambridge University Press)

\bibitem{BE} Erdelyi A et al. 1953
{\it Higher transcendental functions } (NY: McGrow Hill)

\bibitem{Koe} Koekoek R and Swarttouw R F 1994
{\it The Askey-scheme of hypergeometric orthogonal polynomials and its
$q$-analogue}
Report 94-05, Delft University of Technology

\bibitem{KS2} Kuznetsov V B and Sklyanin E K 1996
Separation of variables for $A_2$ Ruijsenaars model  and new integral
representation for $A_2$ Macdonald polynomials
{\it J.\ Phys.\ A: Math.\ Gen.\ }{\bf 29} 2779--2804

\bibitem{Der99} Derkachov S E 1999
Baxter's Q-operator for the homogeneous XXX spin chain
{\it solv-int/9902015, submitted to JPA}

\end{thebibliography}
\end{document}